\documentclass[conference]{IEEEtran}
\IEEEoverridecommandlockouts
\usepackage{cite}
\usepackage{amsmath,amssymb,amsfonts}

\usepackage{algorithmic}
\usepackage{algorithm}
\usepackage{ragged2e}
\usepackage{etoolbox}

\AtBeginEnvironment{algorithmic}{\RaggedRight}

\usepackage{tabularx,makecell,booktabs,array}
\usepackage{graphicx}
\usepackage{caption}
\def\BibTeX{{\rm B\kern-.05em{\sc i\kern-.025em b}\kern-.08em
    T\kern-.1667em\lower.7ex\hbox{E}\kern-.125emX}}

\makeatletter
\def\@texttop{\vskip 8pt}
\makeatother

\begin{document}

\title{Agentic Open RAN: A Deterministic and Auditable Framework for Intent-Driven Radio Control}

\author{
    \IEEEauthorblockN{Hengxu Li\IEEEauthorrefmark{1}, Dongkuan Xu\IEEEauthorrefmark{2}, Mingzhe Chen\IEEEauthorrefmark{3}, Yuchen Liu\IEEEauthorrefmark{2}}

    \IEEEauthorblockA{\IEEEauthorrefmark{1}Tufts University, USA; \IEEEauthorrefmark{2}North Carolina State University, USA; \IEEEauthorrefmark{3}University of Miami, USA}

}

\maketitle

\newcommand{\systemname}{\textsc{A1gent}}

\begin{abstract}
Large language models (LLMs) open new possibilities for agentic control in Open RAN, allowing operators to express intents in natural language while delegating low-level execution to autonomous agents. We present \textbf{\systemname}, an agentic RAN control stack that decouples reasoning from real-time actuation. A non-RT agentic rApp compiles operator goals into typed A1 policy instances, and three task-oriented near-RT agentic xApps enforce them through a deterministic loop with plane-scoped actuation—E2 for mobility and load steering, and O1 for energy orchestration. This agentic reasoning–execution split ensures auditable coordination between RAN intelligent controller (RIC) tiers, supported by encoded guardrails and a fixed-priority action merger for conflict governance. A training-free adaptive policy tuner then refines bounded parameters using KPI memory without retraining, sustaining predictable adaptation. By integrating intent-driven planning with deterministic near-RT execution, \systemname{} advances Open RAN toward verifiable, self-governing, and reproducible agentic intelligence.
\end{abstract}

\section{Introduction}

As mobile networks evolve toward AI-native operation, operators increasingly aim to express \emph{what} the radio access network (RAN) should achieve, rather than micro-managing \emph{how} each control loop behaves. The Open-RAN (O-RAN) architecture, with its layered A1/E2/O1 interfaces, exposes this possibility by separating intent, policy, and actuation. Yet two fundamental gaps remain: (i) how to infuse intent-level reasoning into RAN control without breaking the deterministic guarantees of near-real-time (near-RT) operation, and (ii) how to make such reasoning auditable, replayable, and scoped to well-typed policy knobs~\cite{AIVerification2025, Utkarsh25}.

Specifically, current O-RAN separates intent/policy in the Non-RT domain from actuation over E2/O1 in the near-RT and network domains~\cite{ORANSurvey2025,XAppsSurvey2025,ORANMobility}. Recent studies highlight the need for controllers that can \emph{reason about} operator goals across layers and time scales. Particularly, AgentRAN proposed by~\cite{AgentRAN2025} explicitly argues for a hierarchy of agents decomposing natural-language intents across spatial and temporal scopes, validated over-the-air for power control and uplink scheduling via A1/E2-like exchanges. Yet, its non-deterministic large language model (LLM)-driven decisions raise open questions on stability, multi-agent coordination, and safety-critical guarantees, motivating the elaborated design choice to keep near-RT loops deterministic while bounding LLM reasoning within typed A1 policies. Then, a complementary line of work, Dual-MCP~\cite{DualMCP2025}, proposes a tool-centric, human-in-the-loop planning architecture, where both gNB and user equipment (UE) host Model Context Protocol (MCP) servers exposing tools, resources, and prompts; a cloud LLM planner composes multi-step procedures, validates pre/post-conditions, and logs each action for traceability. This paradigm informs the control decision to preserve auditability and reversible traces between Non-RT reasoning and near-RT actuation.

Meanwhile, \cite{EdgeAgent2025} places an agentic LLM directly \emph{inside} the near-RT RIC, coupled with a long-short term memory (LSTM) predictor to fine-tune power at $\pm1$–$3$\,dB increments. Although effective in simulation, scalability across diverse RAN topologies remains an open issue, underscoring the critical emphasis on keeping near-RT actuation deterministic and LLM-driven adaptation confined to slower, non-RT cycles. Two additional threads, ALLSTaR and AutoRAN, then contribute enabling such capabilities: the former uses LLMs to synthesize MAC schedulers from documentations and package them as distributed (DU)-resident dApps~\cite{ALLSTaR2025}, while the latter leverages Infrastructure-as-Code (IaC) workflows to translate operator intents into continuous deployment pipelines~\cite{AutoRAN2025}. Together, these efforts pave the path toward \emph{Agentic Open RAN}---intent cascading, tool-centric planning, IaC governance, and LLM-assisted synthesis, but none unify them into a deterministic, safety-bound control framework.

To fill this gap, we introduce \textbf{\systemname}, a new control paradigm that reimagines the RIC as an \emph{agentic pipeline} spanning the non-RT and near-RT domains. At its core lies an agentic control stack that decouples reasoning from execution: a non-RT agentic rApp (equipped with LLM and IaC) translates operator goals into \emph{typed A1 policy instances}, while three near-RT agentic xApps (handling quality of service, system load, and energy) enforce them through a fully deterministic control loop. The result is a self-contained, collaborative agentic ecosystem where natural-language intent is compiled into verifiable near-RT behavior, where LLMs plan on second-to-minute horizons and xApps act in milliseconds.

\begin{figure*}[t]
  \centering
  \includegraphics[width=\textwidth]{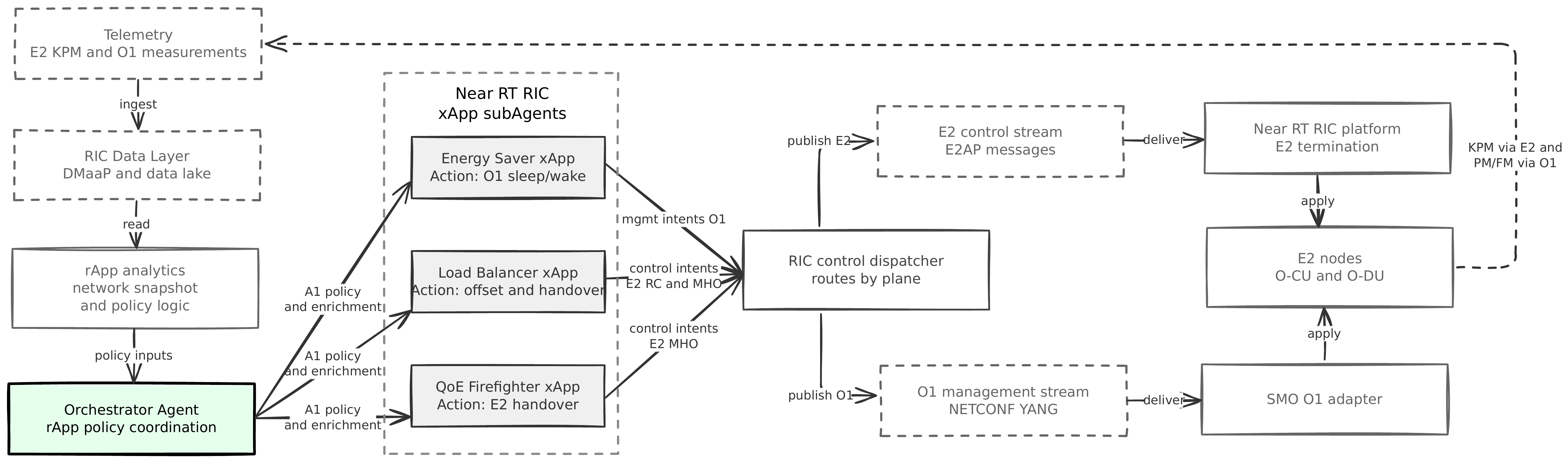}
  \caption{The agentic control loop consists of rApp$\to$A1$\to$xApps, xApps$\to$E2, and SMO/Non‑RT$\to$O1.
  Telemetry (E2–KPM, O1 PM/FM) flows up to the data layer.
  The non‑RT agentic rApp emits typed A1 policy instances to three control components:
  1) QoE (E2SM–MHO), 2) Load (E2SM–RC+MHO), and 3) Energy (O1 via the SMO).
  The dispatcher routes operator intents, while the near-RT RIC executes E2-based control actions and the SMO’s O1 adapter carries out management actions.
  }
  \label{fig:control-plane}
\end{figure*}

\smallskip
\noindent\textbf{Design Overview.}
\systemname{} coordinates multiple task-oriented agents into a coherent architecture within the RIC loop. The non-RT agentic rApp functions as a policy compiler that interprets operator intents, generates typed A1 schemas, and instantiates bounded policy values under strict IaC guardrails, establishing immutable safety envelopes and maintaining versioned audit trails. On the other side, the Near-RT agentic xApps operate within these envelopes through plane-scoped actuation, where the E2 interface handles mobility and load steering, and the O1 interface manages energy orchestration such as sleep and wake operations. A training-free, memory-driven Adaptive Policy Tuner (APT) is designed to adjust only soft policy values at a slow cadence of second-levels. It leverages O-RAN compliant key performance indicator (KPI) summaries and logged action–outcome pairs to refine system behavior while avoiding retraining and preserving explainability. The \textbf{contributions} of this paper are summarized as follows:

\vspace{-0.04cm}
\begin{itemize}
  \item \emph{Agentic reasoning–execution split.} We formalize a typed and auditable coordination between non-RT and near-RT RIC domains: xApps register A1 policy types (\textit{schemas}), while the rApp issues policy instances (\textit{values}). Near-RT loops remain deterministic even without new A1 inputs.

  \item \emph{Safety and conflict governance.} A fixed-priority action merger arbitrates proposals from QoE-aware, system-load, and energy-saving xApps while enforcing IaC-encoded guardrails for both UE and cells.
  \item \emph{Training-free adaptation.} The APT module is designed to leverage rolling KPIs and memory logs to adjust bounded system parameters autonomously, ensuring continuous yet predictable network performance adaptation.

  \item \emph{Integrated E2/O1 orchestration.} QoE-aware and system-load agents operate through E2 (E2SM-MHO/RC), while energy-saving agent acts via O1 through the service management and orchestration (SMO), maintaining reproducibility across control planes.

  \item
  We conduct extensive evaluations using \texttt{ns3-oran} encompassing normal, emergency, and recovery network configuration phases under traffic surge test. The results demonstrate that our \systemname{} effectively elevates network performance while maintaining strict safety guarantees. Specifically, it increases the downlink throughput floors by 20\% and 9\% during the surge phase, sustains a 24\% gain during recovery, and shifts the hotspot load share from 17\% to 44\%. Throughout all phases, the RAN system preserves uplink stability and operates entirely within its agent-governed envelopes.
\end{itemize}

\section{\systemname{}: Agentic Control Stack in Open RAN}
\label{sec:methods}

\subsection{Architecture Overview}
\label{subsec:arch}
As depicted in Fig.~\ref{fig:control-plane}, \systemname{} comprises a non‑RT {orchestrator rApp} and three near‑RT xApps (QoE firefighter, load balancer, energy saver). The agentic rApp ingests operator intent/service level agreement (SLA),  topology, and telemetry summaries, then publishes typed A1 policy instances to each xApp. xApps implement short‑horizon control at 1\,Hz using a minimal metric and action set; the platform routes actuation by plane, i.e. xApps over E2 (QoE via E2SM-MHO; Load via E2SM-RC+MHO), and Energy via O1 (through the SMO) for sleep/wake. Note that the xApps never block on the rApp. When no new A1 policy instance is received, they continue to operate deterministically using the most recently published instance within the enforced guardrails.

\begin{figure*}[t]
  \centering
  \includegraphics[width=\textwidth]{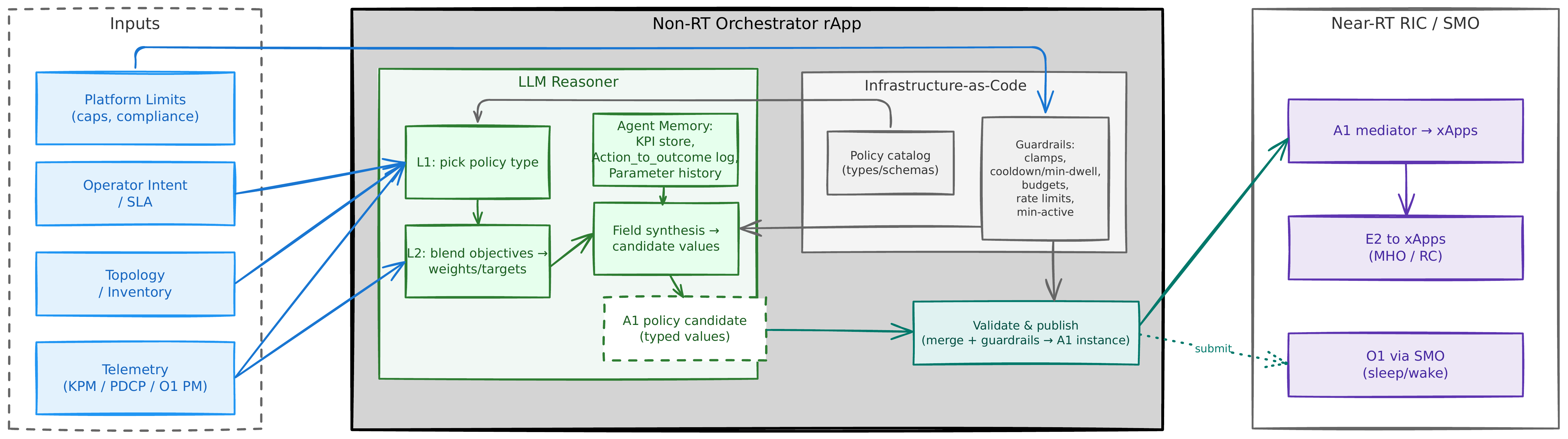}
  \caption{The implementation of agentic control stack with LLM and IaC.
  The LLM side performs intent$\to$A1 translation, Level-1 policy selection, and Level-2 objective blending.
  The IaC side enforces a policy catalog, guardrails, clamps, budgets, and cadence, producing typed A1 policy instances
  for QoE/Load/Energy xApps. A1 carries policy, E2 carries near-RT control, and O1 (via the SMO) carries slow management knobs.
  }
  \label{fig:orchestrator}
\end{figure*}

\subsection{Agent Coordination Mechanisms}
\label{subsec:coord}
The implementation of the agentic control stack is organized into two operational levels driven by the LLM and integrated with IaC, which enforces coordination between the agentic rApp and the xApps, as illustrated in Fig.~\ref{fig:control-plane}.

\paragraph{Level‑1 policy selector (play selection)}
The agentic rApp selects one active play (e.g., QoE‑first, Load‑first, Energy‑first) keyed on phase and coarse telemetry. This activates an A1 policy per xApp with play‑appropriate defaults.

\paragraph{Level‑2 intent translation (turn intent into numbers)}
When objectives compete (e.g., ``protect UE 1 and keep PRB under 85\%''), the rApp converts operator intents into numeric A1 policies, generating per-xApp targets, thresholds, and cadences along with a global budget, interpolated between conservative and aggressive presets within IaC-defined ranges, e.g., translating a handover-trigger offset (a small per-neighbor bias; CIO-style with \( \pm\Delta \,\mathrm{dB} \)) and a downlink (DL) target into:
\[
\begin{aligned}
\texttt{cio\_step\_db}
&=
0.5 + w_{\text{load}}\,(1.5 - 0.5),\\
\texttt{dl\_target\_mbps}
&=
0.3 + w_{\text{qoe}}\,(0.8 - 0.3).
\end{aligned}
\]

The output of this block is a typed A1 instance and near‑RT control loops stay deterministic.

\paragraph{Typed A1 contracts.}
Each xApp registers its A1 policy type (schema), and the rApp publishes policy instances (values). This enforces schema validation and keeps policies auditable.

\paragraph{Conflict management and safety.}

The action merger deduplicates per-UE/per-cell actions and applies a fixed, reproducible priority (e.g., Energy$\!\rightarrow$QoE$\!\rightarrow$Load). A dispatcher then publishes A1 to the xApps through the E2 interface and submits O1 requests to the SMO. Before dispatch, every action is checked against \textit{clamps} (e.g., RC handover-trigger offset $\in[-6,6]$\,dB), \textit{cooldown timers} (per-UE HO bans, per-neighbor offset cooldowns), and budgets (max offset steps per window)~\cite{PACIFISTA2025}.

\subsection{Adaptive Policy Tuner and Memory}
\label{subsec:apt}
The Adaptive Policy Tuner (APT) is a training\mbox{-}free component that proposes small, explainable updates to any A1 policy field as detailed in Table~\ref{tab:a1_fields}. It never blocks near-RT execution and never edits the hard guardrails,
which are enforced by the rApp and xApps on every publication and tick.

The rApp maintains memory of: (i) rolling KPI summaries (per\mbox{-}cell and per\mbox{-}UE distributions for PDCP DL/UL, PRB, SINR/RSRP, MCS, UL interference; per\mbox{-}phase slices); (ii) an action towards the outcome log (offset steps, handovers, sleep/wake, and local effects); and (iii) parameter history for each tunable $\theta$.
On a 30--60\,s cadence, the APT inspects recent KPIs and the `action$\rightarrow$outcome log' to detect gentle drift, then proposes bounded edits to soft tunables. Typical examples are:
\begin{itemize}
  \item \textbf{QoE:} If a cell shows persistent DL shortfalls while the last increase of \texttt{headroom\_min} reduced rescue opportunities with no tail improvement, decrease \texttt{headroom\_min} slightly; if handover (HO) ping\mbox{-}pong rises, increase \texttt{min\_dwell\_s}.
  \item \textbf{Load:} If overload persists despite multiple safe offset steps, increase \texttt{cio\_step\_db} by a small increment; if HO churn spikes, relax \texttt{hot/cool} marks or lower aggressiveness.
  \item \textbf{Energy:} If sleep decisions on the cell frequently auto\mbox{-}revert (i.e. wakes shortly after sleep), tighten \texttt{idle\_prb\_max} or increase \texttt{sleep\_window\_s}; if long idle tails remain active, lower \texttt{idle\_prb\_max} within bounds.
\end{itemize}
Note that these APT’s edits are {small}, {rate\mbox{-}limited}, and {auditable}; they are grounded in simple sign/direction tests from recent history, not in a learned reward.
Each proposal is clamped to IaC ranges, limited by per\mbox{-}parameter max\mbox{-}step and cooldown, optionally EMA\mbox{-}smoothed, and may carry a TTL (auto\mbox{-}revert). All accepted changes are logged with {old}$\rightarrow${new}, {source}, {reason}, and {time}.

\subsection{Governance and Interfaces}
\label{subsec:gov}
A1 carries only {tunable policy fields} as shown in Table~\ref{tab:a1_fields}.
The rApp and its APT may adjust these values at a slow cadence, within admissible ranges.
Hard safety invariants are governed outside A1 and enforced in the rApp prior to publication and at actuation time. They include range limits, minimum dwell, per-cell/UE cooldowns, active sectors, and global budgets. Near-RT xApps enforce these guards locally on every tick.

\begin{table*}[t]
\vspace*{8pt}
\centering
\caption{Defined A1 policy fields (i.e. what travels on A1), with lineage, consumer, and actuator plane.}
\label{tab:a1_fields}
\setlength{\tabcolsep}{7pt}
\renewcommand{\arraystretch}{1.05}
\begin{tabular}{l l l l l l}
\toprule
\textbf{Field} & \textbf{Agent} & \textbf{Kind} & \textbf{Set by (lineage)} & \textbf{Consumed by} & \textbf{Actuator}\\
\midrule
\texttt{DL\_target\_Mbps}     & QoE    & Threshold & Intent $\rightarrow$ L2/LLM       & QoE xApp loop & E2 (MHO) \\
\texttt{SINR\_target\_dB}     & QoE    & Threshold & Intent/Platform $\rightarrow$ L2/LLM & QoE xApp loop & E2 (MHO) \\
\texttt{headroom\_min}        & QoE    & Threshold & APT (LLM) within IaC bounds       & QoE xApp loop & E2 (MHO) \\
\texttt{min\_dwell\_s}        & QoE    & Guard     & APT (LLM) within IaC bounds       & QoE xApp loop & E2 (MHO) \\
\texttt{ue\_ban\_s}         & QoE,\,Load & Cooldown & APT (LLM) within IaC bounds & QoE/Load HO guards; rApp merger & E2 (MHO) \\
\midrule
\texttt{hot\_prb}             & Load     & Threshold & Intent $\rightarrow$ L2/LLM       & Load xApp loop & E2 (RC) \\
\texttt{cool\_prb}            & Load     & Threshold & APT (LLM) within IaC bounds       & Load xApp loop & E2 (RC) \\
\texttt{cio\_step\_db}        & Load     & Offset step & APT (LLM) within IaC bounds       & Load xApp loop & E2 (RC) \\
\texttt{mcs\_min}             & Load     & Guard     & APT (LLM) within IaC bounds       & Load guard     & E2 (MHO) \\
\texttt{ul\_p95\_dBm\_max}    & Load     & Guard     & Platform (non‑LLM)                & Load guard     & E2 (MHO) \\
\midrule
\texttt{idle\_window\_min}        & Energy& Window    & APT (LLM) within IaC bounds       & Energy xApp loop & O1 (SMO) \\
\texttt{idle\_prb\_max}           & Energy& Threshold & APT (LLM) within IaC bounds       & Energy xApp loop & O1 \\
\texttt{idle\_sched\_Mbps\_max}   & Energy& Threshold & APT (LLM) within IaC bounds       & Energy xApp loop & O1 \\
\texttt{idle\_ue\_max}            & Energy& Threshold & Intent $\rightarrow$ L2/LLM       & Energy xApp loop & O1 \\
\texttt{wake\_ue\_min}            & Energy& Threshold & APT (LLM) within IaC bounds       & Energy xApp loop & O1 \\
\texttt{wake\_sched\_Mbps\_min}   & Energy& Threshold & APT (LLM) within IaC bounds       & Energy xApp loop & O1 \\
\texttt{ho\_arrival\_max\_Hz}     & Energy& Guard     & APT (LLM) within IaC bounds       & Energy guard     & O1 \\
\bottomrule
\end{tabular}
\\[4pt]
\raggedright
\end{table*}

\section{Agent-Orchestrated Mobility, Balancing, and Energy Saving}
\label{sec:usecase}

Our agentic control stack is realized through one non-RT rApp and three near-RT xApps that together form a hierarchical, closed-loop architecture. Each component operates at its own cadence but shares typed A1 schemas, KPI logs, and IaC-defined guardrails to ensure deterministic and auditable coordination.

As shown in Algorithm 1, the LLM-assisted orchestrator runs at a 1 Hz tick rate and maintains rolling KPI views for QoE, Load, and Energy over multiple time scales. It transforms operator intents into typed A1 policies through a two-layer reasoning process: (L1) selects a control phase and context; (L2) generates policy values via an LLM prompt constrained by IaC ranges. When the prompt times out, a heuristic blend provides a fallback. The rApp enforces clamp and rate limits, publishes schema-checked A1 policies, merges xApp feedback, and dispatches E2/O1 control batches under global budgets, clamps, and dwell timers.

Algorithm 2 describes the QoE Firefighter agent that enhances user experience by managing handovers. It evaluates neighboring cells with sufficient PRB headroom and uses an LLM prompt (non-blocking) to assess candidate targets. When PDCP or SINR metrics fall below thresholds, and dwell-time constraints are met, the xApp proposes E2-level handover intents to restore QoE. In parallel,
the load-balancing agent (in Algorithm 3) optimizes resource distribution using RC handover-trigger offset updates. It identifies PRB gaps between hot and cool cells, invokes an LLM prompt for context-aware ordering, and applies step-wise offset updates governed by the typed parameter \texttt{cio\_step\_db}. Such stable merge ordering and selective handovers ensure smooth load redistribution while preventing oscillations. Lastly,
the energy-saver agent manages O1-level sleep and wake operations. As described in Algorithm 4, it monitors idle and active cells using PRB, scheduling, and UE activity indicators. When cells remain idle beyond defined windows, the agent issues sleep intents; conversely, when UE activity or scheduled throughput surpass wake thresholds, it proposes wake intents. Both actions are validated through non-blocking LLM prompts and filtered by cooldown and cluster constraints.

\begin{algorithm}[t]
\footnotesize
\caption{Orchestrator rApp (Non-RT): L1/L2 via LLM $\rightarrow$ typed A1; merge; dispatch}
\label{alg:orchestrator-exp}
\begin{algorithmic}[1]
\STATE \textbf{Tick:} 1\,Hz \quad \textbf{State:} last typed A1 (QoE/Load/Energy, versions); KPI rolls; action$\rightarrow$outcome log
\LOOP
  \STATE Build rolling views: QoE (3--10\,s), Load (10--60\,s), Energy (5--10\,min)
  \STATE L1: choose phase; pack goal+context
  \STATE L2 (LLM prompt, deadline $\tau_{\text{LLM}}$): $\widehat{\theta}\!\leftarrow\!\texttt{LLM\_PROMPT}(\text{goal},\text{playbook},\text{views},\text{memory})$; fallback $\texttt{HeuristicBlend}$ if timeout/invalid
  \STATE Clamp/rate-limit: $\theta\!\leftarrow\!\texttt{ClampRate}(\widehat{\theta},\text{IaC})$; publish \textbf{typed A1} (schema-checked, version++)
  \STATE Collect xApp proposals; \textbf{merge} with priority Energy $\triangleright$ QoE $\triangleright$ Load; de-duplicate by \{UE, cell\}
  \STATE Enforce budgets/clamps/dwell/min-active; dispatch E2 batch; submit O1 batch; log
\ENDLOOP
\end{algorithmic}
\end{algorithm}

\begin{algorithm}[t]
\footnotesize
\caption{QoE Firefighter xApp (near-RT; non-blocking LLM {prompt}; E2 HO proposals)}
\label{alg:qoe-exp}
\begin{algorithmic}[1]
\STATE {A1 fields:} \texttt{DL\_target\_Mbps}, \texttt{SINR\_target\_dB}, \texttt{min\_dwell\_s}, \texttt{headroom\_min}
\LOOP
  \FOR{each UE $u$}
    \STATE $C\!\leftarrow$ neighbors with PRB headroom $\ge$\texttt{headroom\_min}; $n_{\text{base}}\!\leftarrow$ best by SINR/PRB
    \STATE $p\!\leftarrow\!\texttt{LLM\_PROMPT\_QoE}(u,C,\text{A1},\text{local KPIs};\ \tau_{\text{xapp}})$ \quad // non-blocking
    \STATE $n^\star\!\leftarrow\!(p.\texttt{preferred\_neighbor}$ if $p$ valid else $n_{\text{base}})$
    \IF{PDCP\_DL($u$)$<$\texttt{DL\_target\_Mbps} \textbf{or} SINR($u$)$<$\texttt{SINR\_target\_dB}}
      \IF{$n^\star$ exists \textbf{and} dwell($u$)$\ge$\texttt{min\_dwell\_s}} \STATE Propose HO($u\!\rightarrow\!n^\star$) \ENDIF
    \ENDIF
  \ENDFOR
\ENDLOOP
\end{algorithmic}
\end{algorithm}

\begin{algorithm}[t]
\footnotesize
\caption{Load Balancer xApp (near-RT; non-blocking LLM {prompt}; E2 RC handover-trigger offset + targeted MHO)}
\label{alg:load-exp}
\begin{algorithmic}[1]
\STATE {A1 fields:} \texttt{hot\_prb}, \texttt{cool\_prb}, \texttt{cio\_step\_db}, \texttt{mcs\_min}, \texttt{ul\_p95\_dBm\_max}
\LOOP
  \STATE $\mathcal{P}\!\leftarrow\!\{(A\!\rightarrow\!B): \text{PRB\_DL}(A)\!\ge\!\texttt{hot\_prb},\ \text{PRB\_DL}(B)\!\le\!\texttt{cool\_prb}\}$
  \STATE $r\!\leftarrow\!\texttt{LLM\_PROMPT\_Load}(\mathcal{P},\text{A1},\text{local KPIs};\ \tau_{\text{xapp}})$ \quad // non-blocking
  \STATE Order $\mathcal{P}$ by PRB gap; if $r$ valid then \texttt{StableMergeOrder}$(\mathcal{P},r.\texttt{pair\_order})$
  \FOR{top $(A\!\rightarrow\!B)\in\mathcal{P}$}
    \STATE Propose RC\_Offset\_Update($A\!\rightarrow\!B$, step $=\pm\,\texttt{cio\_step\_db}$)
    \STATE $e_{\text{base}}\!\leftarrow$ top-PRB UE in $A$ with median SINR$\ge$0\,dB;\quad $e\!\leftarrow\!(r.\texttt{elephant}(A)$ if $r$ valid else $e_{\text{base}})$
    \IF{MCS\_p50($B$)$\ge$\texttt{mcs\_min} \textbf{and} UL\_p95($B$)$\le$\texttt{ul\_p95\_dBm\_max}} \STATE Propose HO($e\!\rightarrow\!B$) \ENDIF
  \ENDFOR
\ENDLOOP
\end{algorithmic}
\end{algorithm}

\section{Evaluation Results}
\label{subsec:problem}

In this section, we evaluate how \systemname{} handles a realistic urban traffic surge in Open RAN without violating safety or cadences. The operator expresses {phase-aware} intent: during \textit{Normal}, keep the network balanced; during \textit{Emergency}, protect vulnerable users (QoE) while preventing hotspots; during \textit{Recovery}, unwind temporary bias and re-enable energy saving.

\begin{algorithm}[t]
\footnotesize
\caption{Energy Saver xApp (near-RT; non-blocking LLM {prompt}; O1 sleep/wake intents)}
\label{alg:energy-exp}
\begin{algorithmic}[1]
\STATE {A1 fields:} \texttt{idle\_window\_min}, \texttt{idle\_prb\_max}, \texttt{idle\_sched\_Mbps\_max}, \texttt{idle\_ue\_max}, \texttt{ho\_arrival\_max\_Hz}, \texttt{wake\_ue\_min}, \texttt{wake\_sched\_Mbps\_min}
\LOOP
  \FOR{each active cell $C$}
    \STATE $idle(C)\!\leftarrow$ windowed tests on PRB, sched DL, UE count, HO arrivals
    \STATE $p\!\leftarrow\!\texttt{LLM\_PROMPT\_Energy}(C,\text{A1},\text{local KPIs};\ \tau_{\text{xapp}})$ \quad // non-blocking
    \IF{$idle(C)$ \textbf{or} ($p$ valid \textbf{and} $p.\texttt{sleep}(C)$)} \STATE Propose \textbf{SLEEP\_INTENT}($C$) \ENDIF
  \ENDFOR
  \FOR{each sleeping cell $C$}
    \STATE $wake\_cond \leftarrow$ UE$\ge$\texttt{wake\_ue\_min} \textbf{or} sched DL$\ge$\texttt{wake\_sched\_Mbps\_min} \textbf{or} cluster PRB high
    \STATE $w\!\leftarrow\!\texttt{LLM\_PROMPT\_Energy}(C,\text{A1},\text{local KPIs};\ \tau_{\text{xapp}})$
    \IF{$wake\_cond$ \textbf{or} ($w$ valid \textbf{and} $w.\texttt{wake}(C)$)} \STATE Propose \textbf{WAKE\_INTENT}($C$) \ENDIF
  \ENDFOR
\ENDLOOP
\end{algorithmic}
\end{algorithm}

\subsection{Scenario Settings}
\label{subsec:scenario}

\textbf{Network topology and traffic mix:} We consider a 3-sector, 3-site macro grid (\(9\) cells) with X2 connectivity using
an extended
\texttt{ns3-oran} simulator~\cite{NISTns3oran2023}. LTE/LENA defaults are adopted with a PF scheduler and enhanced link budget (eNB \(46\)~dBm, UE \(23\)~dBm).
We extend \texttt{ns3-oran} with additional telemetry reporters and control-plane hooks to expose per-cell and per-UE KPIs required for agentic control and evaluation. A total of 20 UEs are deployed in the macro grid. Two thirds follow a random-direction mobility model within an extended rectangular area surrounding the grid, while one third follow a vehicular mobility model with constant velocity and periodic direction reversal to induce macro-border handovers. Radio link failure–based reselection is disabled, so HOs are not overridden by RLF recovery. Each UE runs concurrent traffic flows including eMBB downlink (DL) bursts, URLLC-like short downlink packets (1\,ms cadence), V2X uplink (UL) telemetry (UDP), and sparse mMTC uplink bursts.

\textbf{Baseline Scheme:}
The baseline is an Event~A2 on serving\mbox{-}cell RSRQ (240 ms) plus Event~A4 neighbor RSRQ collection (480 ms); a HO is triggered when the best neighbor exceeds the serving cell by an offset.
We set \texttt{ServingCellThreshold}=28 and \texttt{neighborCellOffset}=1.

\textbf{\systemname{}:}
Native eNB HO (A2/A4-RSRQ) remains enabled so that small RC handover-trigger offset steps can bias HO decisions.
On top, the Load xApp (i) adjusts those offsets via E2SM-RC to steer attach bias and (ii) issues targeted E2SM-MHO handovers for distressed UEs. The rApp deconflicts with A2/A4 using the guards of §\ref{subsec:gov}: per-UE HO ban and dwell timers, per-cell offset intervals and clamps, and suppression of xApp HOs if a recent native HO fired or a cell is under offset cooldown. Sleep/wake is orchestrated over O1 after pre-offload. The LLM used for intent translation and xApp decision proposals is Gemini~2.5~Flash.

\subsection{Experimental Results}
\label{subsec:results}

\paragraph*{1) Emergency tail protection}

\begin{figure}[t]
  \centering
  \includegraphics[width=\linewidth]{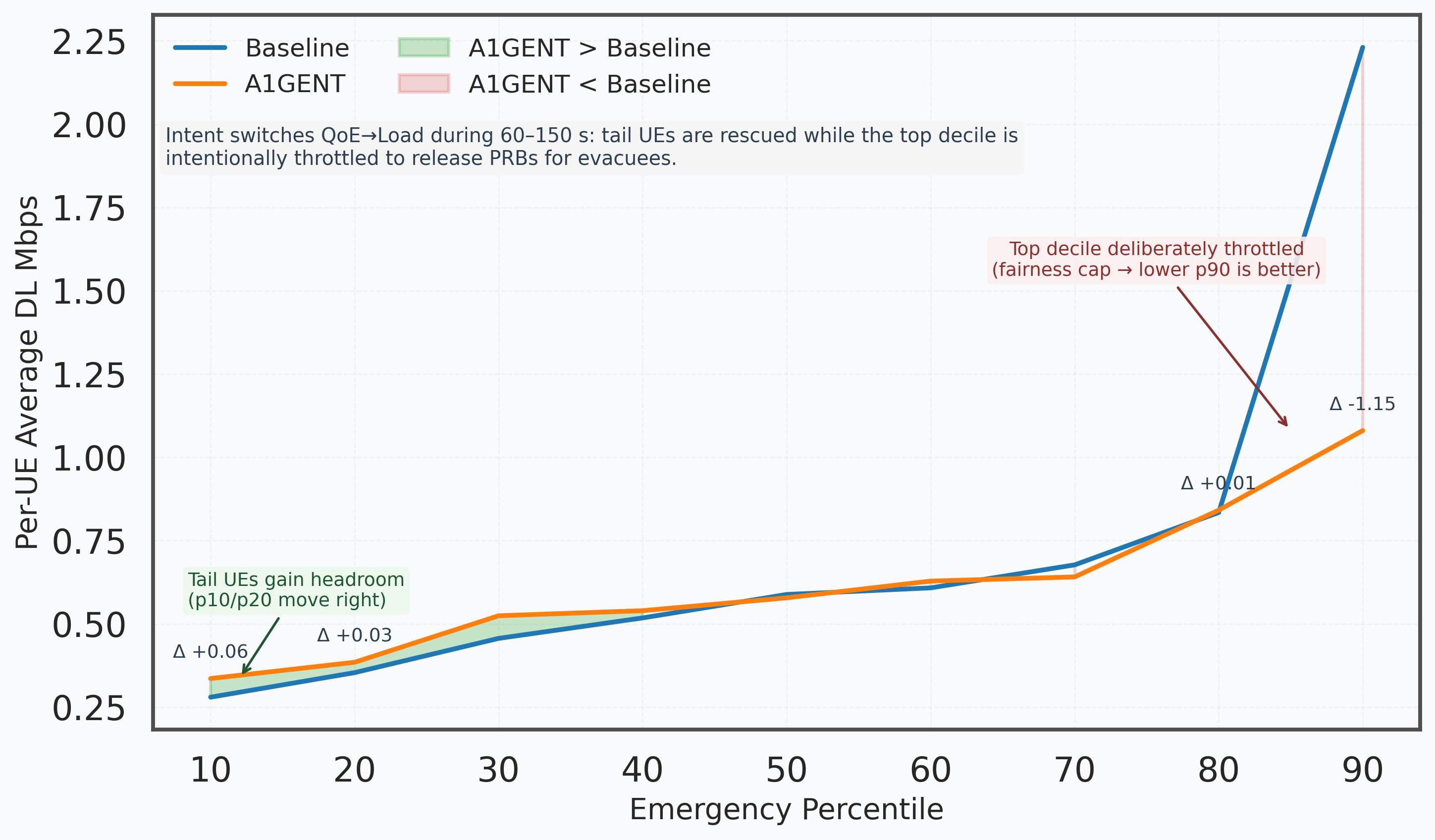}
  \caption{Emergency phase percentile performance. }
  \label{fig:percentile-fan}
\end{figure}

Fig.~\ref{fig:percentile-fan} illustrates the emergency-phase distribution shift, where the weakest UEs gain throughput while the top decile is intentionally reduced to reallocate radio resources. Specifically, per-UE PDCP downlink throughput at p10 increases from \(0.280\) to \(0.336\)\,Mbps and p90 decreases from \(2.230\) to \(1.080\)\,Mbps, reflecting upper-tail compression as a fairness trade-off. Similarly, p10 SINR improves from \(0.34\) to \(0.84\)\,dB under \systemname{}. Table~\ref{tab:tail-fairness} further quantifies tail robustness and emergency fairness.

\begin{table}[!t]
\centering
\caption{Tail robustness (0–300 s) and emergency fairness.}
\label{tab:tail-fairness}
\setlength{\tabcolsep}{4pt}
\renewcommand{\arraystretch}{1.05}
\begin{tabular}{l r r r}
\toprule
\textbf{Metric} & \textbf{Baseline} & \textbf{\systemname{}} & \(\Delta\) \\
\midrule
All-UE DL p05 (Mbps) & 0.000 & \textbf{0.101} & +0.101 \\
Frac. samples $<$ 0.10 Mbps (outage) & 6.4\% & \textbf{4.9\%} & $-1.5$ pp \\
Frac. samples $<$ 0.50 Mbps & 44.4\% & \textbf{39.6\%} & $-4.8$ pp \\
Emergency p90/p10 ratio & 7.96 & \textbf{3.21} & $-59.7$\% \\
\bottomrule
\end{tabular}
\end{table}

\begin{figure}[t]
  \centering
  \includegraphics[width=\linewidth]{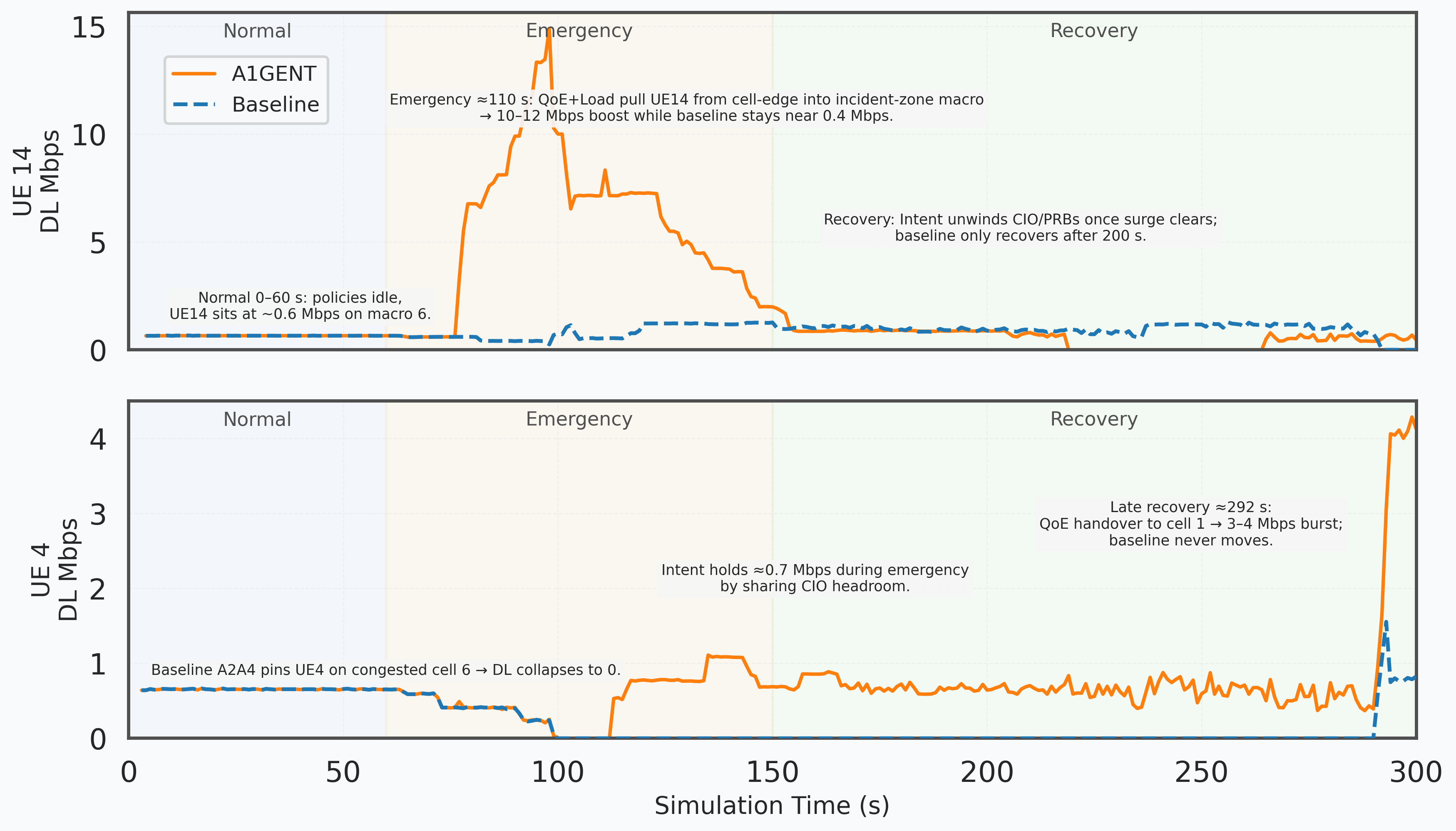}
  \caption{Affected UE timeline across phases.}
  \label{fig:key-ue}
\end{figure}

Fig.~\ref{fig:key-ue} shows the two worst-case scenarios. UE\,14 (cell-edge) jumps from under \(0.2\) to \(10\)–\(12\) Mbps within seconds of the surge and stays above \(5\) Mbps until recovery, driven by targeted E2SM-MHO and higher-quality neighbors.
UE\,4 (incident-zone) improves from \(0.19\) to \(0.53\) Mbps during the surge and reaches \(\sim 4\) Mbps late in recovery, confirming that the benefits persist through our \systemname{}'s adaptive control.

\paragraph*{2) Recovery: sustained QoE and spatial rebalance}
During the recovery phase, user experience continues to improve as network balance is restored. The 20th-percentile downlink throughput increases from \(0.298\) to \(0.370\)~Mbps, representing a 24\% gain. The proportion of samples below \(0.45\)~Mbps decreases from 46.7\% to 43.8\%, and the extreme low-rate tail (\(<0.10\)~Mbps) shrinks from 9.9\% to 6.5\%.
Table~\ref{tab:cell-sched} summarizes mean and median values, showing the incident-zone share of total DL throughput rising from 17\% to \textbf{44\%}. Meanwhile, uplink wideband interference  remains within \(\pm1\)~dB of baseline levels, and the 95th-percentile packet loss remains effectively unchanged at approximately 0.93.

\setlength{\textfloatsep}{5pt}
\begin{table}[!t]
\centering
\caption{Recovery DL scheduler throughput (Mbps).}
\label{tab:cell-sched}
\setlength{\tabcolsep}{4pt}
\renewcommand{\arraystretch}{1.05}
\begin{tabularx}{\columnwidth}{c r r >{\raggedleft\arraybackslash}X}
\toprule
\textbf{Focus Cell} & \textbf{Baseline mean} & \textbf{Ours mean} &
\makecell[r]{\textbf{Median:} \textbf{Baseline$\to$Ours}}\\
\midrule
1 & 1.62 & \textbf{2.52} & 0.00$\to$\textbf{2.47} \\
2 & 3.78 & 4.19         & 3.09$\to$3.40 \\
3 & 3.64 & 3.66         & 3.09$\to$3.09 \\
9 & 2.29 & \textbf{4.45} & 0.00$\to$\textbf{2.47} \\
\bottomrule
\end{tabularx}
\end{table}

\begin{figure}[t]
  \centering
  \includegraphics[width=0.86\linewidth]{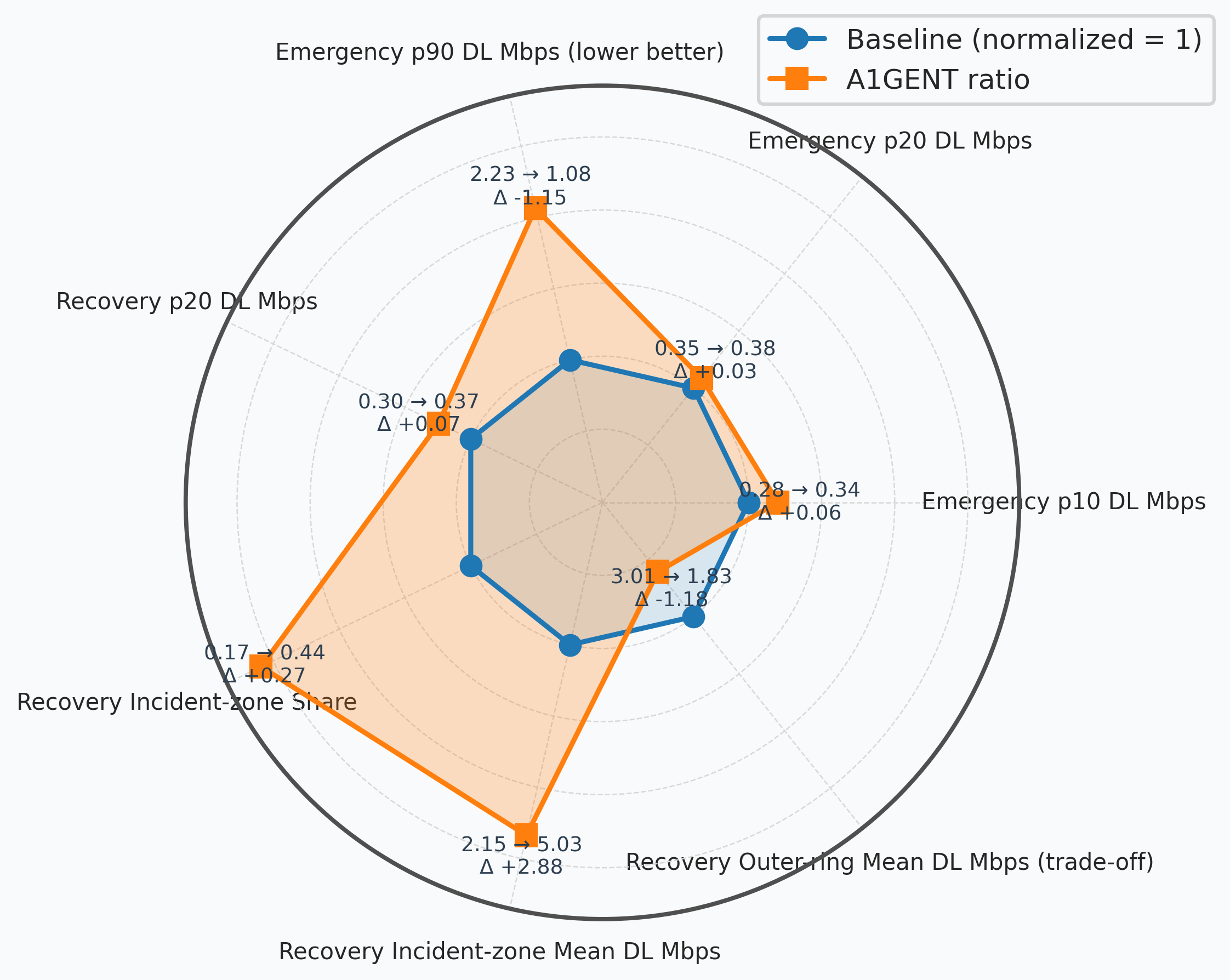}
  \caption{Capability radar summarizing \systemname{} performance. }
  \label{fig:capability-radar}
\end{figure}

\paragraph*{3) Mobility and HO stability}
Traffic directionality becomes structured and goal-driven under our agentic policy. In the baseline, most mobility events are concentrated toward cell~6, whereas \systemname{} exhibits balanced bidirectional flows (1$\leftrightarrow$6 and 3$\leftrightarrow$6), reflecting targeted offloading during the surge phase and controlled repatriation afterward. Dwell-time statistics further confirm improved mobility stability, where the 90th-percentile dwell time during the emergency phase increases from 43\,s to 53\,s, indicating fewer ping-pong HOs, while the 99th-percentile dwell time during recovery remains stable (189.6$\rightarrow$188.0\,s).
Fig.~\ref{fig:capability-radar} summarizes the key performance gains and trade-offs, comparing \systemname{} against the baseline.

\section{Conclusions}
\label{sec:conclusions}
We presented \systemname, an agentic control stack that keeps near‑RT execution deterministic and auditable while carrying operator intent through typed A1 policy instances in Open RANs. The non‑RT orchestrator based on LLM and IaC translates intent into value‑only A1 instances, while near‑RT xApps execute fixed‑rate loops with IaC‑enforced guardrails. Future works will focus on integrating hardware-in-the-loop capabilities for E2/O1 control and expanding the A1 policy catalog to cover additional O-RAN control domains.

\vspace{0.42cm}
\section{Acknowledgment}

This research was supported by NSF through Award CNS--2440756, CNS--2312138, CNS--2332834, and NVIDIA Academic Grant Program.

\vspace{12pt}


\vspace{0.43cm}
\begin{thebibliography}{00}

\bibitem{AIVerification2025}
R. Soundrarajan, C. Fiandrino, M. Polese, S. D'Oro, L. Bonati, and T. Melodia,
``On AI Verification in Open RAN,''
\textit{IEEE Communications Magazine}, early access, 2025,
doi: 10.1109/MCOM.001.2500318.

\bibitem{Utkarsh25}
U. Sharma, H. Wei, M. Chen, J. Xu, and Y. Liu,
``Adaptive traffic steering in Open RAN: Integrating rule-based policies with reinforcement learning,''
in \textit{Proc. IEEE INFOCOM Workshops}, 2025.

\bibitem{ORANSurvey2025}
K. Alam et al.,
``A comprehensive tutorial and survey of O-RAN: Exploring slicing-aware architecture, deployment options, use cases, and challenges,''
\textit{IEEE Commun. Surv. Tutor.}, vol. 28, pp. 1637--1678, 2026, doi: 10.1109/COMST.2025.3598406.

\bibitem{XAppsSurvey2025}
A. Elyasi, A. Ashdown, K. M. Rumman, and F. Restuccia, ``O-RAN xApps: Survey and research challenges,'' \textit{Comput. Netw.}, 2025.

\bibitem{ORANMobility}
M. Mushi et al.,
``Open RAN testbeds with controlled air mobility,''
\textit{Computer Communications},
vol. 228, p. 107955, 2024,
doi: 10.1016/j.comcom.2024.107955.

\bibitem{AgentRAN2025}
M. Elkael et al.,
``AgentRAN: An agentic AI architecture for autonomous control of open 6G networks,''
arXiv:2508.17778 [cs.AI], 2025.

\bibitem{DualMCP2025}
E. Baena, A. Mandal, and D. Koutsonikolas, ``Demo: human-in-the-loop agentic reconfiguration of edge 5G networks via dual-MCP and LLM reasoning,'' in \textit{Proc. ACM MobiHoc}, Houston, TX, USA, 2025, pp. 391--392, doi: 10.1145/3704413.3765314.

\bibitem{EdgeAgent2025}
A. Salama, Z. Nezami, M. M. H. Qazzaz, M. Hafeez, and S. A. R. Zaidi,
``Edge agentic AI framework for autonomous network optimisation in O-RAN,''
in \textit{Proc. IEEE Int. Symp. Pers., Indoor and Mobile Radio Commun. (PIMRC) Workshops}, 2025.

\bibitem{ALLSTaR2025}
M. Elkael, M. Polese, R. Prasad, S. Maxenti, and T. Melodia,
``ALLSTaR: Automated LLM-driven scheduler generation and testing for intent-based RAN,''
arXiv:2505.18389 [cs.NI], 2025.

\bibitem{AutoRAN2025}
S. Maxenti et al.,
``AutoRAN: Automated and zero-touch open RAN systems,''
arXiv:2504.11233 [cs.NI], 2025.

\bibitem{PACIFISTA2025}
P. Brach del Prever et al., ``PACIFISTA: Conflict evaluation and management in Open RAN,'' \textit{IEEE Trans. Mobile Comput.}, vol. 24, no. 10, pp. 10590--10605, 2025, doi: 10.1109/TMC.2025.3570632.

\bibitem{NISTns3oran2023}
W. Garey, R. A. Rouil, E. Black, T. Ropitault, and W. Gao, ``O-RAN with machine learning in ns-3,'' in \textit{Proc. Workshop on ns-3 (WNS3)}, Ballston, VA, USA, 2023, pp. 60--68, doi: 10.1145/3592149.3592157.

\end{thebibliography}
\end{document}